\documentclass[prd, twocolumn, superscriptaddress]{revtex4}

\usepackage{epsfig}
\usepackage{graphicx}
\usepackage{amsmath}
\usepackage{amsfonts}
\usepackage{epstopdf}

\def\slashchar#1{\setbox0=\hbox{$#1$}     		
   \dimen0=\wd0                                 	
   \setbox1=\hbox{/} \dimen1=\wd1               	
   \ifdim\dimen0>\dimen1                        	
      \rlap{\hbox to \dimen0{\hfil/\hfil}}      	
      #1                                        	
   \else                                        	
      \rlap{\hbox to \dimen1{\hfil$#1$\hfil}}   	
      /                                         	
   \fi}

\begin{document}

\title{The Chiral Magnetic Effect}

\author{Kenji Fukushima}
\email{fuku@yukawa.kyoto-u.ac.jp}
\affiliation{Yukawa Institute, Kyoto University, Kyoto, Japan}

\author{Dmitri E. Kharzeev}
\email{kharzeev@bnl.gov}
\affiliation{Department of Physics,
Brookhaven National Laboratory, Upton NY 11973, USA}

\author{Harmen J. Warringa}
\email{warringa@quark.phy.bnl.gov}
\affiliation{Department of Physics,
Brookhaven National Laboratory, Upton NY 11973, USA}

\date{\today}

\begin{abstract}
  Topological charge changing transitions can induce chirality in the
  quark-gluon plasma by the axial anomaly.  We study the equilibrium
  response of the quark-gluon plasma in such a situation to an
  external magnetic field.  To mimic the effect of the topological
  charge changing transitions we will introduce a chiral chemical
  potential.  We will show that an electromagnetic current is
  generated along the magnetic field.  This is the Chiral Magnetic
  Effect.  We compute the magnitude of this current as a function of
  magnetic field, chirality, temperature, and baryon chemical
  potential.
\end{abstract}

\maketitle

\section{Introduction}
The quark-gluon plasma is a phase of extremely hot matter consisting
of quarks and gluons. Just after the Big-Bang, the universe itself
was in the quark-gluon plasma phase. The quark-gluon plasma can be created
and studied using collisions of heavy ions. An active experimental
program to investigate the properties of this hot phase of matter is
underway using the Relativistic Heavy Ion Collider (RHIC) at BNL.\ \ 
In the near future the quark-gluon plasma will also be studied using
the Large Hadron Collider (LHC) at CERN, the Facility for
Antiproton and Ion Research (FAIR) at GSI, and the NICA facility
at JINR, Dubna.

The behavior of the quark-gluon plasma is described by Quantum
Chromodynamics (QCD). One of the intriguing predictions of QCD is that
in the quark-gluon plasma phase certain special gluon configurations
to which one can assign a winding number play a role
\cite{BPST,H}. This winding number is a topological invariant, which
means that smooth deformations of these configurations do not change
the winding number.  Experimental evidence for the existence of
configurations with nonzero winding number is only indirect from the
meson spectrum~\cite{Witten:1979vv,'tHooft:1986nc,Chu:1994vi}.

The configurations with nonzero winding number are in fact transitions
which invoke passing a potential barrier with a height of order 
the QCD scale $\Lambda_{\mathrm{QCD}}$ over the strong coupling constant $\alpha_S$. Because of the height of
the barrier, the transitions are highly suppressed at low temperatures
since they require tunneling~\cite{H}. The configurations responsible
for this tunneling process are called instantons~\cite{H,
  'tHooft:1986nc, GPY, SS98}.  At high temperatures in the quark-gluon
plasma phase, it is possible to jump over the potential barrier. The
transitions are therefore not suppressed anymore and called sphalerons
\cite{M83, KM84, KRS, AM87, KS88, AM88}. These configurations were studied in
the electroweak theory as a mechanism for baryogenesis 
\cite{KRS, AM87, AM88, Shaposhnikov1987}, and are
also relevant for QCD \cite{MMS, GZ94, SZ03}. 

At these high temperatures the configurations with nonzero winding
number can be produced with relatively high probability \cite{MMS,
  BMR}. Therefore the quark-gluon plasma is the best place to find
direct experimental evidence for the existence of gauge field
configurations with nonzero winding number.

These configurations do something very distinct to quarks; they can,
depending on the sign of their winding number, transform left- into
right-handed quarks or vice-versa via the axial anomaly \cite{ABJ}
(see also \cite{C80, S92}).  For massless quarks, the axial anomaly
equates $\partial_\mu j_5^\mu$ to the topological term.  The spatial
integration of $\partial_\mu j_5^\mu$ yields an exact relation for the
rate of the chirality change induced by topological configurations,
which reads
\begin{equation}
 \frac{\mathrm{d} (N_R - N_L) }{\mathrm{d} t}
 = -\frac{g^2 N_f}{16\pi^2} \int \mathrm{d^3}x\, 
  F^{\mu \nu}_a \tilde F^a_{\mu \nu},
\label{eq:axialanomaly}
\end{equation}
where $N_{R,L}$ denotes the net number of quarks (minus antiquarks)
with right/left-handed chirality, $N_f$ the number of massless
flavors, and $\tilde
F^a_{\mu\nu}=\tfrac{1}{2}\epsilon_{\mu\nu\lambda\sigma}F^{\lambda\sigma
  a}$, with $\epsilon^{0123} = 1$.  All the massless flavors equally
couple to the gauge field, hence the proportionality factor $N_f$
arises in Eq.~(\ref{eq:axialanomaly}).  Let us stress that, in the
common convention, chiral quarks have opposite helicity to antiquarks;
a particle with right-handed chirality has right-handed helicity,
while an anti-particle with right-handed chirality has left-handed
helicity.  For instance the helicity of the antineutrino $\bar{\nu}_L$
is right-handed.  Here right-handed helicity means spin and momentum
parallel, while left-handed helicity means spin and momentum
anti-parallel.  Therefore the difference $N_R - N_L$ can also be read
as the total number of quarks plus antiquarks with right-handed
helicity minus the total number of quarks plus antiquarks with
left-handed helicity.  For physical gluon configurations
(configurations with finite action) the time integral over the
right-hand side of Eq.~(\ref{eq:axialanomaly}) is equal to minus twice
the winding number of the gluon field configuration.  As a result of
the axial anomaly the interactions between these configurations and
the quarks break the parity ($\mathcal{P}$) and charge-parity
($\mathcal{CP}$) symmetry.  Ordinary (perturbative) interactions
between quarks and gluons cannot induce a difference between the
number of right- and left-handed quarks.  A mass term always will tend
to wash out such difference \cite{AGP}.

In QCD, the probability to generate either a gluon configuration with
positive or negative winding number is equal. This is because there is
no direct $\mathcal{P}$ and $\mathcal{CP}$ violation in QCD (assuming
the value of the $\theta$ angle is equal to zero). In the quark-gluon
plasma, many of these configurations can be generated at different
points in space and time with different winding numbers.  In pure
$\mathrm{SU}(N)$ Yang-Mills theory this process is completely random; the
dynamics of the chirality change is that of a one-dimensional random
walk. In QCD with massless flavors, however, it will cost energy to
induce a difference between the number of right- and left-handed
quarks because of the Fermi-principle. Therefore the dynamics is not
completely random anymore, and there is a preference to decrease the
chirality \cite{MMS}.  In any case, the variance of the chirality will
be nonzero, and increase as a function of time according to
diffusion. Hence, it is expected that every time the quark-gluon
plasma is produced, it will posses a non-zero chirality \cite{KKV}.
The chirality averaged over many events of quark-gluon plasma
production vanishes.  Therefore one speaks in this case of
event-by-event $\mathcal{P}$- and $\mathcal{CP}$-violation.

Next to the sphaleron transitions, chirality could also be introduced
in the quark-gluon plasma in the same way due to chromoelectric and chromomagnetic fields in the
initial state of the quark-gluon plasma produced in heavy-ion
collisons, i.e.\ the so-called glasma \cite{KKV, LM06,KKL}.
Although the net topological charge cannot develop with the Boost invariant configuration \cite{KKV}, the glasma instability spontaneously breaks the Boost invariance \cite{glasmainst}, so that the event-by-event topological charge fluctuation is expected.  The situation is then quite reminiscent of the sphaleron transitions \cite{LM06}.

It was first argued by one of us \cite{K06} that if $\mathcal{P}$- and
$\mathcal{CP}$-violating processes are taking place in the quark-gluon
plasma produced in heavy-ion collisions, then positive charges should
separate from negative charges along the direction of angular
momentum of the collision.  In Ref.~\cite{KZ} this mechanism was
worked out in more detail using an effective $\theta$ angle to mimic
$\mathcal{P}$- and $\mathcal{CP}$-violating processes.  In heavy-ion
collisions the magnetic field is pointing in the direction of angular
momentum. It was shown in Ref.~\cite{KZ} that this magnetic field
induces charge on a $\theta$-domain-wall in such a way that an
electric field is created parallel to the magnetic field.  In this way
positive charge is separated from negative charge along the magnetic
field.  In Ref.~\cite{KMW} a different mechanism for charge-separation
was discussed (see also \cite{WQM}). It was shown that a magnetic field
in the presence of imbalanced chirality induces a current along the
magnetic field. Again, as a result, positive charge is separated from
negative charge along the magnetic field.  This is called the ``Chiral
Magnetic Effect''.

Due to the separation of charge along the direction of the magnetic
field in heavy-ion collisions, an asymmetry between the amount of
positive/negative charge above and below the reaction plane is
expected \cite{K06, KZ, KMW}.  These asymmetries can be analyzed in
experiments using a correlation study as proposed by Voloshin
\cite{V04}. Preliminary data from the STAR collaboration is presented
in Refs.~\cite{IVS, VQM}.  Observation of the Chiral Magnetic Effect
will be direct experimental evidence for the existence of
topologically non-trivial gluon configurations. It furthermore is
evidence for event-by-event $\mathcal{P}$- and
$\mathcal{CP}$-violation.

The Chiral Magnetic Effect could be used to determine whether a
deconfined chirally symmetric phase of matter is created in heavy-ion
collisions \cite{KMW}. Deconfinement is a necessary requirement for
the Chiral Magnetic Effect to work, since it requires that
soft quarks can separate over distances much greater than the 
radius of the nucleon. Moreover, chiral symmetry restoration is essential,
because a chiral condensate will tend to erase any asymmetry between
the number of right- and left-handed fermions.

In this article we will investigate the Chiral Magnetic Effect in
detail. In order to treat the non-vanishing chirality, we introduce a
chiral chemical potential, denoted as $\mu_5$.  The chiral chemical
potential will be generated by the topological charge changing
transitions. We will not study this dynamical process, but we just
assume this chemical potential is there.  Then we will study the
implications of applying a magnetic field to a system with nonzero
chiral chemical potential in equilibrium.  We will see that an
electromagnetic current will be induced in the direction of the
magnetic field. We will compute the magnitude of this current as a
function of magnetic field, chirality, temperature, and baryon
chemical potential.

Besides the Chiral Magnetic Effect, the fact that a magnetic field can influence QCD processes is well
known. For example a magnetic field can induce chiral symmetry
breaking \cite{GMS}, influence the chiral condensate \cite{SS97, CMW},
and therefore modify the phase diagram of QCD (see Refs.~\cite{FM, AF}
for recent discussions). Also the color-superconducting phases
predicted to exist at high baryon densities are strongly affected by
a strong magnetic field \cite{ABR, FIM, FW, NS}. Finally the anomaly
in the presence of a magnetic field can give rise to all kinds of
interesting effects, like spontaneous creation of axial currents
\cite{SZ04, MZ} and formation of $\pi^0$-domain walls \cite{SS08}.

The analysis we present in this article can be used to make
predictions for the charge asymmetries in heavy-ion collisions like are
done in Ref.~\cite{KMW}. We will encounter the beautiful physics of
the anomaly, current quantization and the index theorem, and periodic
oscillatory behavior due to Landau level quantization.

\section{Chiral chemical potential}
\begin{figure}[t]
\includegraphics{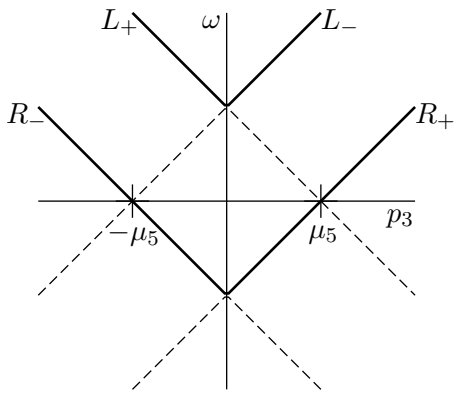}
\caption{Spectrum of massless Dirac fermions with right- and
  left-handed chirality in the presence of an chiral chemical
  potential $\mu_5$.  The subscript $\pm$ denotes the eigenvalue of
  the spin in the $z$-direction. The chiral chemical potential induces
  a nonzero density of right-handed particles and left-handed anti-particles.
}
\label{fig:diracspec}
\end{figure}

As was argued in the introduction, topological charge changing
transitions can induce an asymmetry between the number of right- and
left-handed quarks due to the axial anomaly.  In order to study the
effect of this asymmetry we introduce a chiral chemical potential
$\mu_5$. This chemical potential couples to the difference between the
number of right- and left-handed fermions. To the Lagrangian
density the following term is added
\begin{equation}
 \mu_5 \bar \psi \gamma^0 \gamma^5 \psi.
\label{eq:chiralchempot}
\end{equation}
The energy spectrum of the free Dirac equation in the presence of a
chiral chemical potential is for massless modes (with $p_x=p_y=0$ for
simplicity),
\begin{eqnarray}
 \omega_{R\pm} = \pm p_3 - \mu_5, \\
 \omega_{L\pm} = \mp p_3 + \mu_5. 
\end{eqnarray}
Here $\pm$ represents the spin in the $z$-direction and $R$, $L$ the
chirality.  The momentum in the $z$-direction is given by $p_3$; let
us stress that in our notation $p_3$ does not denote the third
component of a four-vector with metric convention.  We have displayed
the massless energy spectrum in Fig.~\ref{fig:diracspec}. In the
massless limit one can distinguish modes with right-handed chirality
from modes with left-handed chirality.  It should be mentioned that
$p_3$ is restricted to be positive for the $R+$ and $L-$ particle
modes so that the helicity is positive for $R+$ and negative for $L-$,
respectively, and $p_3$ is negative for the $R-$ and $L+$ particle
modes (see Fig.~\ref{fig:diracspec}).  If the chiral chemical
potential is positive some of the right-handed particle modes will
become occupied while some of the left-handed anti-particle modes will
be filled as well. A net chirality is created in this way.

The chiral chemical potential lifts the degeneracy between modes with
right- and left-handed chirality. A difference between the total
number of particles plus anti-particles with right-handed and
left-handed helicity is created. The magnetic field will lift the
degeneracy in spin depending on the charge of the particle. Hence
particles with right-handed helicity will tend to move opposite to
anti-particles with right-handed helicity. As a result an
electromagnetic current is generated along the magnetic field, which
is the Chiral Magnetic Effect \cite{KMW} (see also Refs.~\cite{KMW,
  WQM} for a pictorial representation of the Chiral Magnetic Effect).
We will compute this induced electromagnetic current in the next
section.

The effect of a finite amount of topological charge change can also be
mimicked by an effective theta angle, which could depend on space-time
(see for example \cite{K06, KZ, KPT}). 
One adds to the Lagrangian of QCD the following term, 
\begin{equation}
 \frac{g^2}{32\pi^2} \theta(x,t) F^{\mu \nu}_a \tilde F^{a}_{\mu \nu}.
\end{equation}
By performing an axial U(1) rotation this term can
be transformed into the following fermionic contribution
\begin{equation}
 \frac{1}{2 N_f} \partial_\mu \theta \,\bar \psi \gamma^\mu \gamma^5 \psi.
\end{equation}
Identifying this with Eq.~(\ref{eq:chiralchempot}) we see that $\mu_5
= \partial_0 \theta / 2 N_f$. We can also identify $\mu_5$ with the time
component of an axial vector field $A_\mu^5$.
The effective theta angle results in a difference between the rates of
changing left-handed into right-handed and changing right-handed into
left-handed particles. The chiral chemical potential, however, is a
more static quantity; it is the energy necessary to put a right-handed
quark on its Fermi surface or to remove a left-handed quark from its
Fermi surface.  It describes the difference between the number of
right- and left-handed fermions. An effective theta angle to describe
spontaneous $\mathcal{P}$ and $\mathcal{CP}$-violating processes has
been discussed often in the literature (for examples see
Refs.~\cite{K06, KZ, KPT, theta}).  The chiral chemical potential has
on the other hand only been used in a few papers \cite{NM, MMS, SSS98,
  JPT}.

Let us finally point out that the chiral chemical potential has no
sign problem, i.e. the fermionic determinant with $\mu_5$ is real and
positive.  In the presence of a chiral chemical potential the
fermionic determinant reads in Euclidean space-time,
\begin{equation}
 \mathrm{det} \mathcal{M}(\mu_5)
 \equiv
 \mathrm{det} \left( \slashchar{D} + \mu_5 
 \gamma^0_E \gamma^5  + m \right),
\end{equation}
where $\slashchar{D}= \gamma_E^\mu D_\mu$. Here we have chosen a
representation in which all $\gamma_E$ matrices are Hermitian,
$\gamma^0_E = \gamma^0, \gamma^i_E = i \gamma^i$.  Since
$\slashchar{D}$ and $\gamma_E^0 \gamma^5$ are anti-Hermitian the
eigenvalues of $\mathcal{M}(\mu_5)$ are of the form $i \lambda_n + m$,
where $\lambda_n \in \mathbb{R}$. Because $\gamma_5$ anticommutes with
$\slashchar{D} + \mu_5 \gamma_E^0 \gamma^5$, all eigenvalues come in
pairs, which means that if $i \lambda_n + m$ is an eigenvalue, also
$-i \lambda_n + m$ is an eigenvalue. Since the determinant is the
product of all eigenvalues we see that the determinant is the product
over all $n$ of $\lambda_n^2 + m^2$.  Hence the determinant is real
and also positive semi-definite. This is very interesting because it
allows for a lattice QCD simulation of chirally asymmetric systems.
The lattice QCD can then simulate the Chiral Magnetic Effect by
introducing a space-dependent phase on the link variable which amounts
to the external magnetic field.

\section{Computation of Induced Current}
In this section we will show if a magnetic field is applied to a
system with an asymmetry between the number of right- and left-handed
fermions, an electric current is induced along the magnetic field.  We
will compute this current in four different ways, since we think they
are all very instructive.  The first way is through an energy balance
argument. Then we will arrive at the result by solving the Dirac
equation. The third way is by explicitly computing the thermodynamic
potential in the presence of a magnetic field.  The last derivation we
discuss is using a derivative expansion of the effective action.  We
will compute the current for a generic fermion with electric
charge $e$ and neither flavor nor color; at the end of this section we
will discuss what happens if we recover the flavor and color for
quarks.

Let us set up notation.  We will take the metric $g_{\mu \nu} =
\mathrm{diag}(1, -1, -1, -1)$ and the chiral representation for the
gamma matrices;
\begin{equation}
 \gamma^\mu = 
\left(
\begin{array}{cc} 
0 & \sigma^\mu \\ \bar \sigma^\mu & 0
\end{array}
\right), 
\;\;\;
 \gamma^5 = 
\left(
\begin{array}{cc} 
-1 & 0 \\ 0 & 1
\end{array}
\right),
\end{equation}
where $\sigma^\mu = (1, \sigma^i)$ and $\bar \sigma^\mu = ( 1,
-\sigma^i)$ are the quaternion bases.  Using this convention it is
possible to write the fermion field $\psi$ into its left- and
right-handed components $\psi = (\phi_L, \phi_R)^T$.  We define the
right- and left-handed chemical potentials as $\mu_R = \mu + \mu_5$
and $\mu_L = \mu - \mu_5$.  Here $\mu$ denotes the quark chemical
potential, which for three colors ($N_c=3$) is equal to one third
  of the baryon chemical potential.  If we write $p_3$, as we
mentioned in the previous section, we mean the $z$-component of the
momentum vector $\vec p$ and not the third component of the
four-vector $p_\mu$.

The total current is equal to the volume integral over the current
density,
\begin{equation}
 J^\mu = \int \mathrm{d}^3 x\, j^\mu(x).
\label{eq:thetotalcurrent}
\end{equation}
The current density is given by the following expectation value,
\begin{equation}
 j^\mu(x) = e \langle \bar \psi (x)  \gamma^\mu \psi(x) \rangle. 
\end{equation}
Here the expectation value is over a thermodynamic ensemble. One can
write the current density in terms of right- and left-handed
spinors as
\begin{equation}
 j^\mu(x) = e \langle \phi_R^\dagger(x) \sigma^\mu \phi_R(x) \rangle 
+ e \langle \phi_L^\dagger(x) \bar \sigma^\mu \phi_L(x) \rangle.
 \label{eq:thecurdens}
\end{equation}

\subsection{Axial anomaly and the energy balance} 
The easiest way to obtain the right expression for the current is
using a beautiful argument of energy balance by Nielsen and Ninomiya
\cite{NM}.  Consider a situation with an electric field
$\boldsymbol{E}$ and a magnetic field $\boldsymbol{B}$ in the presence
of a chiral chemical potential.  In that case the electromagnetic
anomaly will tell us that the rate of change of chirality is equal
to the volume integral over $e^2 \boldsymbol{E} \cdot \boldsymbol{B}
/ 2 \pi^2$.  There exists an intuitive derivation of this rate
  \cite{NM} which we will repeat here.

Let us consider fermions with positive charge $e$ in a
background magnetic field $\boldsymbol{B}$. The fermions will occupy
Landau levels, so their motion in the transverse (to the field
$\boldsymbol{B}$) plane will be restricted. The fermions however are
free to move along or opposite the direction of $\boldsymbol{B}$;
since the spins of the fermions are preferentially aligned along the
field, the motion parallel to $\boldsymbol{B}$ corresponds to the
right-handed fermions, and anti-parallel to $\boldsymbol{B}$ -- to 
left-handed fermions.

The presence of an electric field $\boldsymbol{E}$ parallel to
$\boldsymbol{B}$ causes the chirality to change (see
Ref.~\cite{Witten:1984eb} for a related discussion of particle
acceleration in cosmic strings). The energy of right-handed
fermions moving along the electric field under the influence of the
Lorentz force will grow linearly with time, leading to the growing
Fermi momentum,
\begin{equation}\label{fermi_mom}
p_F^R = e E t.
\end{equation}
Likewise, for left-handed charges the Fermi momentum will decrease,
with $p_F^L = - p_F^R$; this corresponds to the production of
left-handed anti-particles with charge $-e$.  Therefore the particles
with charge $e$ will move along the field, and anti-particles with
charge $-e$, against the field. Thus, an electric current is created
along $\boldsymbol{E}$.

The density of right-handed fermion states is equal to the product of
the longitudinal phase space density $\mathrm{d}n/\mathrm{d}z =
p_F^R/2\pi$ and the density of Landau levels in the transverse
direction $\mathrm{d}^2n/\mathrm{d}x\, \mathrm{d}y = eB/2\pi$,
\begin{equation}
\frac{p_F^R}{2 \pi} \cdot \frac{e B}{2 \pi} = 
\frac{e^2}{4 \pi^2}\ \boldsymbol{E} \cdot \boldsymbol{B}\ t .
\end{equation}
The same expression yields also the density of left-handed
anti-fermion states; therefore, the rate of chirality $N_5 = N_R -
N_L$ generation per unit volume per unit time is then given by
\begin{equation}
\frac{\mathrm{d}^4 N_5}{\mathrm{d}t\,\mathrm{d}^3x} = 
 \frac{e^2}{2 \pi^2}  \ \boldsymbol{E} \cdot \boldsymbol{B}.
\end{equation}
We have thus reproduced the general anomaly relation for the
electromagnetic fields.

Consider now the energy balance related to the chirality change.  To
change a left-handed fermion in a right-handed fermion requires
removing a particle from the left-handed Fermi surface and adding it
to the right-handed Fermi-surface.  This will cost an energy $\mu_R -
\mu_L = 2 \mu_5$ or $\mu_5\,\mathrm{d}N_5$. So multiplying this
energy by the rate of chirality change we know how much energy is
needed per unit of time. This energy has to come from somewhere,
assuming no losses; it will be equal to the power delivered by a
current.  This power is equal to the product of the current with the
electric field. So one finds \cite{NM}
\begin{equation}
 \int \mathrm{d}^3 x \, \boldsymbol{j} \cdot \boldsymbol{E} = 
 \mu_5 \frac{\mathrm{d}N_5}{\mathrm{d}t} =
 \frac{ e^2 \mu_5}{2\pi^2}  \int \mathrm{d}^3 x \,\boldsymbol{E}
 \cdot \boldsymbol{B}.
\end{equation}
We can take $\boldsymbol{E}$ in the direction of $\boldsymbol{B}$ in this
expression. Then if we take the limit $\boldsymbol{E} \rightarrow 0$ we find
\begin{equation}
  \boldsymbol{J}  = 
 \frac{ e^2 \mu_5}{2\pi^2}
\int \mathrm{d}^3 x \, \boldsymbol{B}.
\label{eq:currentenergycons} 
\end{equation}
This derivation clearly shows that not only the axial anomaly of
QCD plays a role in the Chiral Magnetic Effect, but also the
electromagnetic axial anomaly. The QCD anomaly provides the chirality,
the electromagnetic anomaly the current.
In a box with periodic boundary conditions, the number
of Landau levels is an integer. This gives rise to current
quantization as we will closely see in the next microscopic
derivation.

\subsection{Dirac equation}
We will now compute the induced current by solving the Dirac equation
in the presence of a magnetic field and chiral chemical potential.  We
take the magnetic field in the $z$-direction,
\begin{equation}
 \boldsymbol{B} = B(x, y)\, \mathbf{e}_z.
 \label{eq:magfield}
\end{equation}
The Dirac equation in this background reads
\begin{equation}
 \left( i \gamma^\mu \mathcal{D}_\mu - m + 
\mu \gamma^0 + \mu_5 \gamma^0 \gamma^5 
\right) \psi(x) = 0.
\end{equation}
where $D_\mu = \partial_\mu - i e A_\mu$. In order to incorporate the
magnetic field given in Eq.~(\ref{eq:magfield}) the only non-vanishing
components of $A_\mu$ are $\mu = 1, 2$. Furthermore $A_\mu$ only will
depend on $x$ and $y$. The precise form of $A_\mu$ is not relevant for
our calculation.

We will compute the total current in the $z$-direction as is given
in Eq.~(\ref{eq:thetotalcurrent}) starting from Eq.~(\ref{eq:thecurdens}).
To proceed one has to make a momentum decomposition of the fields in
terms of creation and annihilation operators.  As is shown explicitly
in Ref.~\cite{MZ} the only non-vanishing contribution to
\begin{equation}
\int  \mathrm{d}^3 x\, \langle \phi_{R,L}^\dagger(x) \sigma^3 \phi_{R,L}(x)
\rangle
\end{equation}
arises from the transverse zero modes, i.e. modes which have
$p_x=p_y=0$.  The reason is that in all the non-zero modes there is a
spin degeneracy in energy, which results in the cancellation of the
expectation value of $\sigma_3$ \cite{AC, MZ}.  The transverse
zero-modes are however not degenerate.  Let us denote the number of
transverse zero modes with $\sigma_3$ equal to $\pm$ as $N_\pm$. One
shows that the difference $N_+ - N_-$ is equal to the index of a
two-dimensional Dirac Hamiltonian in the presence of a magnetic field
\cite{AC}. This index can be expressed in terms of the total flux
$\Phi$. One finds \cite{AC, MZ}
\begin{equation}
N_+ - N_- = \Bigl \lfloor \frac{e}{2\pi}\Phi \Bigr \rfloor, 
\end{equation}
where we have introduced the floor function $\lfloor x \rfloor$ which is the
largest integer smaller than $x$.  The flux is equal to the integral
of the magnetic field over the transverse plane,
\begin{equation}
\Phi =  \int \mathrm{d}^2 x\, B(x,y).
\end{equation}
Let us stress here that the number of zero modes is quantized,
and not the magnetic flux itself.

It is now possible to construct the total current. It is equal to the
sum of number densities in the transverse zero-mode weighted by the
spin degeneracy $\pm N_\pm$. 
For the right-handed modes we find
\begin{multline}
\int \mathrm{d}^3 x 
\langle \phi_{R\pm}^\dagger \sigma_3 \phi_{R\pm} \rangle
= 
\pm N_\pm L_z \int_0^{\infty} \frac{\mathrm{d}p_3}{2 \pi}
\Bigl[n( p_3 - \mu_R)\\
-
n( p_3 + \mu_R) \Bigr ]
= \pm N_\pm \frac{L_z \mu_R}{2\pi}.
\label{eq:righthandedcontribution}
\end{multline}
Here $L_z$ denotes the length of the system in the $z$-direction and
$n(\omega) = [\exp(\omega/T) + 1]^{-1}$ is the Fermi-Dirac
distribution. The two Fermi-Dirac distributions in
Eq.~(\ref{eq:righthandedcontribution}) correspond to right-handed
particle and antiparticle modes respectively.  In front of the
antiparticle contribution there is a minus sign, since $\phi^\dagger
\phi$ is the number density of particles minus antiparticles. The
temperature dependence has dropped out from
Eq.~(\ref{eq:righthandedcontribution}) without approximation.
The reason why $p_3$ runs only positive is, as we have explained on
Fig.~\ref{fig:diracspec}, $R+$ has positive $p_3$ only and $R-$ has
negative $p_3$ whose sign we changed in the integral. Similarly, for
the left-handed modes we find,
\begin{multline}
\int \mathrm{d}^3 x\, 
\langle \phi_{L\pm}^\dagger \sigma_3 \phi_{L\pm} \rangle
= 
\pm N_\pm L_z \int_0^{\infty} \frac{\mathrm{d}p_3}{2 \pi}
\Bigl[n( p_3 - \mu_L)\\
-
n( p_3 + \mu_L) \Bigr ]
= \pm N_\pm \frac{L_z \mu_L}{2\pi}.
\end{multline}

By taking the spin sum and subtracting L from R contributions we find
that the total current becomes
\begin{equation}
 J = e 
\Bigl \lfloor
\frac{e \Phi}{2\pi} \Bigr \rfloor \frac{L_z \mu_5}{\pi} .
\label{eq:totalcurrent}
\end{equation}
The result is independent of temperature and density.  By adding the
two contributions one finds the total induced axial current, $J_5 =
\int \mathrm{d}^3 x\, \langle \bar \psi \gamma^3 \gamma^5 \psi
\rangle$ in the massless limit,
\begin{equation}
 J_5 =  
\Bigl \lfloor
\frac{e \Phi}{2\pi} 
\Bigr \rfloor
\frac{L_z \mu}{\pi} .
\end{equation}
This current was computed for $\mu_5=0$ by Metlitski and Zhitnitsky
\cite{MZ}.  We recover the result of Ref.\ \cite{MZ} and find that the
total axial current is independent of $\mu_5$.

This derivation can be performed in the more general case with massive
fermions. The computation is more involved, but the final answer will
turn out to be independent of mass.  In the next derivation we will
include a mass term and show that the answer is independent of
mass. There, it will be transparent why the result is insensitive to
temperature and density as well. The last derivation using the
derivative expansion will provide understanding why the current is
independent of mass from a different point of view.

\subsection{Thermodynamic potential}
We will now derive the current in a homogeneous magnetic background
using the thermodynamic potential.  
In the presence of a chiral chemical potential we
find that the thermodynamic potential is given by
\begin{multline}
 \Omega = \frac{\vert eB \vert}{2\pi} \sum_{s=\pm} \sum_{n=0}^{\infty}
\alpha_{n,s}
\int_{-\infty}^{\infty} \frac{\mathrm{d} p_3}{2\pi}
\Bigl [ \omega_{p, s} 
\\
+ T \sum_\pm \log (1 + e^{-\beta (\omega_{p, s} \pm \mu) })
\Bigr ],
\label{eq:thermopotmag}
\end{multline}
where $n$ is a sum over Landau levels, $s$ is a sum over spin and the
dispersion relation is given by
\begin{equation}
 \omega_{p,s}^2 = \left[
\mathrm{sgn}(p_3)  
(p_3^2 + 2 \vert  e B \vert n)^{1/2} +
s \mu_5  \right]^2 + m^2.
\label{eq:thermopotmagdispers}
\end{equation}
The first term in the square brackets may also be written as
$p_3(1+2|eB|n/p_3^2)^{1/2}$ without the sign function.  The constant
$\alpha_{n,s}$ ensures that the lowest Landau level only contains one
spin component,
\begin{equation}
\alpha_{n, s} = \left \{ 
 \begin{array}{cl} 1  & n > 0,
 \\ \delta_{s+}&  n = 0, \;\;eB > 0,
\\ \delta_{s-}&  n = 0, \;\;eB < 0.
\end{array} \right. 
\end{equation}
We also note again that the phase space associated with Landau levels
is quantized in a box with periodic boundary conditions.
We omit this to avoid bothersome notation like
$\lfloor eL_xL_yB/2\pi\rfloor/L_xL_y$ in the phase space factor.

Let us introduce a constant gauge field $A_3$. One might think that a
constant gauge field could be gauged away, but this is not possible by
a gauge transformation satisfying the periodic boundary condition.
The current density is the derivative of the thermodynamic potential
with respect to $A_3$ at the point $A_3=0$,
\begin{equation}
 j_3 = \left. \frac{\partial \Omega}{\partial A_3}
\right \vert_{A_3=0}.
\end{equation}
The thermodynamic potential is still given by
Eq.~(\ref{eq:thermopotmag}), but the dispersion relation
Eq.~(\ref{eq:thermopotmagdispers}) is now modified by replacing $p_3$
by $p_3 + e A_3$.  In order to regularize the ultraviolet divergences
of thermodynamic potential we introduce a momentum cutoff $\Lambda$ on
the $p_3$ integral. Furthermore we introduce a cutoff $N$ on the sum
over the Landau levels.  After we have introduced this regularization
we can pull the derivative with respect to $A_3$ through the sum and
integral. Then we can use that
\begin{equation}
 \frac{\partial}{\partial A_3} =
 e \frac{\mathrm{d}}{\mathrm{d} p_3},
\end{equation}
when acting on an arbitrary function of $\omega_{p,s}$.
As a result we find the following expression
for the current density,
\begin{multline}
 j_3 = e \frac{\vert e B \vert}{2\pi} \sum_{s=\pm} \sum_{n=0}^{N}
\alpha_{n,s}
\int_{-\Lambda}^{\Lambda} \frac{\mathrm{d} p_3}{2\pi}\,
\frac{\mathrm{d}}{\mathrm{d} p_3}
\Bigl [ \omega_{p,s} \\
+ T\sum_\pm \log(1+e^{-\beta(\omega_{p, s} \pm \mu)})
\Bigr ],
\end{multline}
where $\omega_{p,s}$ is now given by
Eq.~(\ref{eq:thermopotmagdispers}) since we used that $A_3$ has to put
to zero after taking the derivative. After summing over spins the
contribution to the integrand of the Landau Levels with $n>0$ is an
odd function of $p_3$. Hence only the lowest Landau level which
contains one spin component contributes to the current. As a result
for $eB > 0$ we find
\begin{equation}
 j_3 = e \frac{\vert e B \vert}{2\pi} 
\int_{-\Lambda}^{\Lambda} \frac{\mathrm{d} p_3}{2\pi}
\frac{\mathrm{d}}{\mathrm{d} p_3}
\Bigl [ \omega_{p,+} + T\sum_\pm \log(1+e^{-\beta(\omega_{p, +} \pm \mu)})
\Bigr ],
\label{eq:currentthermopot}
\end{equation}
where
\begin{equation}
 \omega_{p,\pm}^2 = (p_3 \pm \mu_5 )^2 +m^2.
\end{equation}
For $e B < 0$ one has to replace $\omega_{p,+}$ by $\omega_{p,-}$ in
Eq.~(\ref{eq:currentthermopot}).  Since the integrand is a total
derivative, it is easily integrated.  The medium part (logarithmic
term) drops because it goes to zero with $p_3\to\pm\infty$.  Only a
surface term remains, which equals
\begin{align}
 j_3 &= e \frac{\vert e B \vert}{4\pi^2}
\left[ 
\omega_{p,\pm}(p_3 = \Lambda)
-
\omega_{p,\pm}(p_3 = -\Lambda)
\right]
\nonumber \\
&  = e\frac{\vert e B\vert}{4\pi^2}\left[
  (\Lambda\pm\mu_5) - (\Lambda\mp\mu_5)\right] = 
\frac{e^2 \mu_5}{2\pi^2} B,
\label{eq:curdensthermo}
\end{align}
where we have used that $\pm$ corresponds to the sign of $eB$.
The fact that the current is equal to a surface term
is because it is caused by the electromagnetic
anomaly. This as was argued in the first derivation.

By multiplying the current density Eq.~(\ref{eq:curdensthermo}) with
the volume one finds the total current Eq.~(\ref{eq:totalcurrent}).
The virtue in this derivation is that it is manifest that the current
results from the surface integral at infinitely large momentum, to
which any infrared effects of mass, temperature, and $\mu$ are
irrelevant.  The next derivation using the derivative expansion will
give us more understanding why this result is independent of mass.

\subsection{Derivative expansion of effective action}
The last derivation of the current we discuss is by using a derivative
expansion of the effective action as is performed by D'Hoker and Goldstone
\cite{HG} (see also \cite{BO}). Let us introduce an axial vector field
$A_\mu^5$ and write the covariant derivative as $\mathcal{D}_\mu
= \partial_\mu - ieA_\mu -i e A_\mu^5 \gamma^5$. One can define right-
and left-handed vector fields as follows: $A_{R} = A_\mu + A_\mu^5$
and $A_{L} = A_\mu - A_\mu^5$.  By performing the integration over the
fermions fields one obtains the following effective action
\begin{equation}
  S_{\mathrm{eff}} = \mathrm{log}\, \mathrm{Det} 
  \left( i \slashchar{D} - m \right).
\end{equation}
Here $\mathrm{Det}$ includes the space-time coordinates as well as the
color and Dirac indices. The current density $j^\mu$ can be obtained
by taking the functional derivative of this expression with respect to
$A_\mu$. In the presence of an axial vector field the divergence of a
vector current is anomalous; one has \cite{HG}
\begin{equation}
 \partial_\mu j^\mu = e \frac{e^2}{16\pi^2} \left(
F^{\mu \nu}_L \tilde F^{\phantom{\mu}}_{L,{\mu \nu}}
-
F^{\mu \nu}_R \tilde F^{\phantom{\mu}}_{R,{\mu \nu}}
\right).
\label{eq:vectoranomaly}
\end{equation}
One can write down an expansion of the current in terms of the
fields $A_\mu$, $A_\mu^5$ and their derivatives. The expression
should be Lorentz covariant and $\mathrm{U}(1)$ gauge invariant.
Furthermore the current should satisfy the
anomaly constraint Eq.~(\ref{eq:vectoranomaly}).
To first order in the fields and derivatives one obtains \cite{HG},
\begin{equation}
j^{\mu} = -\frac{e^2}{4\pi^2} \epsilon^{\mu \nu \rho \sigma}
   e A_\nu^5 F_{\rho \sigma}.
\label{eq:covcurrent}
\end{equation}
The current is $m$-independent. This follows directly from the
anomalous divergence of the vector current, that has no $m$-dependent
contributions even with inclusion of a mass term.  However, the
divergence of the axial vector current is $m$-dependent.  Therefore
the axial vector current induced by a magnetic field depends on mass.
This is indeed found in Ref.~\cite{MZ}.

We can now use that $e A_0^5 = \mu_5$ in Eq.~(\ref{eq:covcurrent}), 
so that we obtain the current density induced by a magnetic field,
\begin{equation}
\boldsymbol{j} = \frac{e^2 \mu_5}{2\pi^2} \boldsymbol{B}.
\label{eq:dercurrent}
\end{equation}
Since the last equation was obtained via a derivative expansion,
the derivation assumes constant magnetic fields.

\subsection{Discussion of derivations}
We have argued in Sec.~II that $A_\mu^5 = \partial_\mu \theta /2 N_f$
up to a coupling constant.  Suppose we have a space-dependent theta
angle $\theta$, for example formed by a domain wall.  The covariant
current in Eq.~(\ref{eq:covcurrent}) shows that an electric field will
induce a current perpendicular to the electric field on the domain
wall. Moreover, it shows that a magnetic field will induce charge on
the domain wall.  The generation of charge on domain walls or solitons
was first discussed by Goldstone and Wilczek \cite{GW}.  Callan and
Harvey \cite{CH} have studied this mechanism as well in the context of
axionic cosmic strings. They however use pseudoscalar coupling instead
of axial vector coupling, but find a result for the current which is
equivalent to Eq.~(\ref{eq:covcurrent}).  It was argued in
Refs.~\cite{BDF, FDB} that on domain walls formed in certain
semi-conductors currents could be generated perpendicular to the
electric field for the same reason. In the context of charge
separation in heavy-ion collisions, the generation of charge on
$\theta$ domain walls was discussed by Kharzeev and Zhitnitsky
\cite{KZ}.

Goldstone and Wilczek \cite{GW} have derived their current
using a perturbative one-loop calculation. It is also possible
to compute our current perturbatively. One obtains a triangle
one-loop diagram with two vector couplings and one axial vector
coupling. As is well known, this diagram contains the anomaly.
If one includes the effect of the chiral chemical potential
in the fermion propagator, the diagram to compute is the
photon polarization tensor.

The axial anomaly generates the topological term which is a color
singlet.  So no net color is separated by the Chiral Magnetic
Effect. Hence it is expected that no additional chromo-electric fields
are built up along the direction of the magnetic field. Therefore a
possible gluonic back-reaction can be neglected. This can also be
inferred from Eq.~(\ref{eq:vectoranomaly}), since it will not be
modified by the presence of a gluonic background field.  As a result,
the expression for the current Eq.~(\ref{eq:dercurrent}) is correct
even in the presence of a time-independent gluonic field.

If the Chiral Magnetic Effect operates in a heavy-ion collision, the
current is generated in a finite volume. Hence charges are separated,
so an electric field will be built up along the direction of the
magnetic field. This could cause a back-reaction.  We think that in
the study for the implications in heavy-ion collisions, this
back-reaction can be neglected, since the electric field is small
compared to the magnetic field (it only involves a few charges, while
the magnetic field is created by all charges). Furthermore the
electric force is small compared to the gluonic force.

We have obtained the current for one fermion with charge $e$. In the
quark-gluon plasma there are 3 relevant quark flavors, up, down and
strange with charges $q_f = 2/3 e, -1/3e$ and $-1/3e$ which have
$N_c = 3$ colors. The total
current will be the sum of the contributions of the individual ones,
which follow from the previous obtained expressions by replacing $e$
with $q_f$, summing over flavors and multiplying by the number of colors.
This results in 
\begin{equation}
 J = N_c \sum_f q_f \Bigl \lfloor 
\frac{q_f \Phi}{2\pi} 
\Bigr \rfloor 
\frac{L_z \mu_5}{\pi}.
\end{equation}

\section{Current expressed in chiral charge}
As we saw in the previous section, the induced current is proportional
to $\mu_5$.  The chiral chemical potential $\mu_5$ is a
parameter which induces an asymmetry between the number density of
right- and left-handed fermions $n_5=n_R-n_L$. Since the asymmetry is
conserved by varying the magnetic field or the temperature, $\mu_5$
will depend on the magnetic field, temperature, and chemical potential.
In this section we will compute the conserved quantity $n_5$ as a
function of $\mu_5$. We then will express $\mu_5$ in terms of $n_5$ in
order to obtain the dependence of the induced current on $n_5$. This
allows us to make comparisons of the magnitude of the Chiral Magnetic
Effect in different situations. Moreover, it allows us to relate
the current to sphaleron dynamics, since the change in $N_5$ is
equal to $-2N_f$ times the winding number of the sphaleron.

In the computation we present here we will neglect the effect of the
gluons.  At very large temperatures, this is correct, since the
coupling between gluons and quarks is small due to asymptotic
freedom. However, at smaller temperatures the relation between $\mu_5$
and $n_5$ could be modified by gluonic corrections. A perturbative
calculation and/or a lattice simulation could give insight in the
relevance of these corrections.  We leave the computation of gluonic
corrections for future investigation.  For QCD, the results we obtain
at zero temperature are therefore unreliable. However, since we keep
the discussion general, these results could be of use for a system of
non-interacting fermions. Again, we will take $N_c = N_f = 1$ in the
computations. At the end we present the high temperature QCD
result with multiple flavors.

In contrast to the computation of current, it is difficult to perform
the full analytical evaluation of the chiral charge density.  This is
because transverse non-zero modes have contributions unlike the
current which originates from zero modes only.  We will, therefore,
consider two simple limits analytically; the weak and strong
magnetic field cases in order. Outside these limit we will
resort to a numerical calculation.

\subsection{Weak magnetic field limit}
In the weak magnetic field limit ($ \vert e B \vert < \mu_5^2$) we can
expand the current in powers of $ \vert e B \vert / \mu_5^2$.  To
leading order it is enough to compute the total chiral charge density
$n_5 = n_R - n_L$ in the absence of a magnetic field. To compute the
chiral charge density we first construct the thermodynamic potential,
\begin{multline}
 \Omega = \sum_{s=\pm} \int \frac{\mathrm{d}^3 p}{(2\pi)^3}
\Bigl [ \omega_{p, s}
+ T \sum_\pm \log (1 + e^{-\beta (\omega_{p, s} \pm \mu) })
\Bigr ]
\end{multline}
with
\begin{equation}
 \omega_{p,s}^2 = (p + s \mu_5)^2 + m^2, 
\end{equation}
where $p = \vert \vec p \vert $. By differentiating the thermodynamic
potential with respect to $\mu_5$ one finds the chiral charge density
which reads
\begin{equation}
n_5
 =
\frac{1}{2\pi^2} 
\sum_{s=\pm} 
 \int_0^{\infty} \!\!\! \mathrm{d} p\, p^2\,
\frac{\mu_5 + s p }{\omega_{p,s}} 
\left[ 1 - \sum_{\pm} n(\omega_{p,s} \pm \mu) \right].
\end{equation}
In the massless limit we find
\begin{equation}
n_5 = \frac{1}{3\pi^2} \mu_5^3 +
\frac{1}{3} \mu_5 \left (T^2 + \frac{\mu^2}{\pi^2} \right).
\label{eq:n5}
\end{equation}

We can now compute the current in a general magnetic field which is
assumed to be small compared to $\mu_5^2$.  Let us define the
average magnetic field as
\begin{equation}
 \langle B \rangle = \frac{1}{L^2} \Phi ,
\end{equation}
assuming $L_x=L_y=L_z=L$.
Let us for a moment assume that $\Phi \gg 2\pi/e$ such that
to good approximation we can ignore the effects of current
quantization. We will discuss this effect in the next subsection.

For temperatures and chemical potentials smaller $\mu_5$
we find $\mu_5 \approx (3\pi^2)^{1/3} n_5^{1/3}$
such that
\begin{equation}
J = 
 \frac{(3\pi^2)^{1/3}}{2\pi^2}
e^2 L^2 N_5^{1/3} \langle B \rangle  .
\label{eq:zerotempcurrent}
\end{equation}
Let us define the Chiral Magnetic conductivity
$\sigma_B$ as
\begin{equation}
 \sigma_B = \frac{J}{\langle B \rangle}.
\end{equation}
The Chiral Magnetic conductivity now becomes at zero temperature,
\begin{equation}
 \sigma_B =
  \frac{(3\pi^2)^{1/3}}{2\pi^2}
e^2 L^2 N_5^{1/3}.
\label{eq:smalltmu}
\end{equation}

For temperatures and/or quark chemical potentials larger than $\mu_5$
we find $\mu_5 \approx 3 n_5 /(T^2+\mu^2/\pi^2)$. In that case the
current yields
\begin{equation}
 J = \frac{3 e^2}{2 \pi^2}
\frac{1}{T^2 + \mu^2/\pi^2}
  N_5 \langle B \rangle .
\label{eq:hightempcurrent}
\end{equation}
The calculation shows that the Chiral Magnetic conductivity
in the high-temperature limit is
\begin{equation}
  \sigma_B =  \frac{3 e^2}{2 \pi^2}
\frac{1}{T^2 + \mu^2/\pi^2} N_5 .
\label{eq:largetmu}
\end{equation}

\begin{figure}[t]
\includegraphics{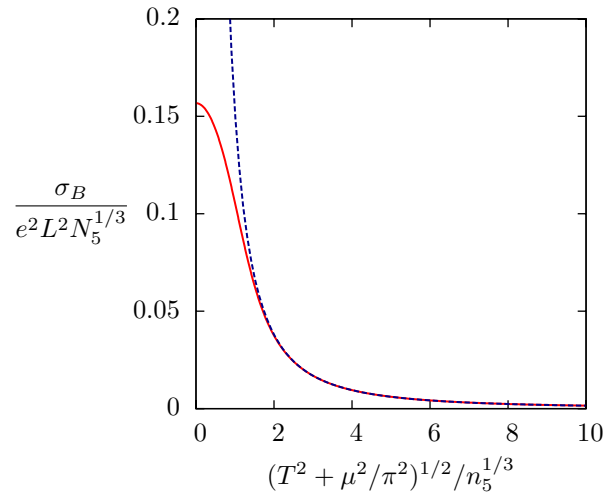}
\caption{Chiral Magnetic conductivity as a function of temperature
and chemical potential.
The dashed line is the high-temperature/chemical
potential approximation.}
\label{fig:conductivity}
\end{figure}

We have displayed the Chiral Magnetic conductivity as a function of
temperature and chemical potential in Fig.~\ref{fig:conductivity}
using Eq.~(\ref{eq:n5}).  The figure shows that $\sigma_B$ begins from
Eq.~(\ref{eq:smalltmu}) and approaches Eq.~(\ref{eq:largetmu}) as
either $T$ or $\mu$ grows.

The reason that the current drops as a function of temperature and
chemical potential is that in both cases $\mu_5$ should take a smaller
value for a given $n_5$ through Eq.~(\ref{eq:n5}).  This is because a
medium at finite $T$ and $\mu$ has fermion distributions with higher
momenta that can take part in the chiral charge density.  Conversely,
a smaller $\mu_5$ is sufficient to mimic the effect of a given $n_5$
at high temperature/chemical potential, that means that the effect of
$n_5$ on systems diminishes by temperature and chemical potential.
The generated current decreases accordingly because it is proportional
to $\mu_5$.

Let us briefly discuss current quantization here.
If the size of the magnetic field is large compared to the area in
which it is confined the effects of current quantization become
important.  For example, consider a magnetic field which is constant
within a tube with radius $R$ and vanishes outside.  The effects of
current quantization become important if 
the total flux $R^2 B$ is comparable to the flux quantum $2\pi/e$.

\begin{figure}[t]
\includegraphics{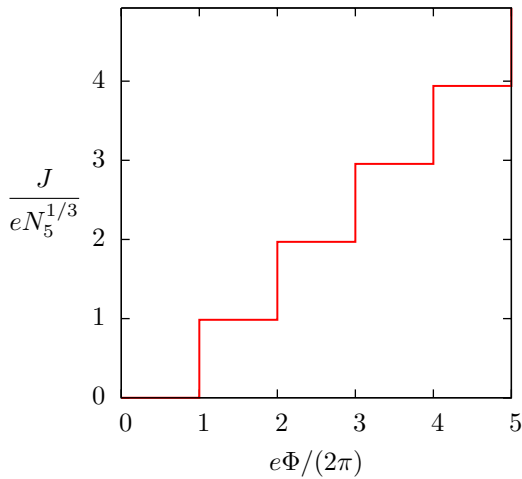}
\caption{Current as a function of flux for $T=0$.}
\label{fig:fluxquant}
\end{figure}

In Fig.~\ref{fig:fluxquant} we have displayed the current as a
function of the flux for $T=\mu=0$.  Clearly one can see the
quantization of the current. If $e \Phi$ becomes an integer multiple
of $2\pi$ another zero mode is available, which result in an increase
of the current by an amount of $e N_5^{1/3} (3\pi^2)^{1/3}/\pi \sim
0.985 e N_5^{1/3}$.

\subsection{Homogeneous magnetic field}
Now let us investigate the current in a strong magnetic field.  We now
will take a homogeneous field in which we can calculate the induced
chiral charge as a function of $\mu_5$.  Again we start from the
thermodynamic potential which in the presence of a homogeneous
background magnetic field and a nonzero $\mu_5$ is given by
Eq.~(\ref{eq:thermopotmag}).  Differentiating the thermodynamic
potential with respect to $\mu_5$ gives the chiral charge,
\begin{multline}
n_5
 = 
\frac{\vert e B \vert}{2\pi} \sum_{s=\pm} \sum_{n=0}^{\infty}
\alpha_{n,s}
\int_{-\infty}^{\infty} \frac{\mathrm{d} p_3}{2\pi}
\frac{\mathrm{d} \omega_{p,s}}{\mathrm{d} \mu_5} \times 
\\
\Bigl [ 1
- \sum_{\pm} n(\omega_{p,s} \pm \mu) \Bigr],
\end{multline}
where
\begin{equation}
\frac{\mathrm{d} \omega_{p,s}}{\mathrm{d} \mu_5}
=
\frac{\mu_5 + \mathrm{sgn}(p_3)
(p_3^2 + 2 \vert  e B \vert n)^{1/2} s
}{\omega_{p,s}}.
\end{equation}

In the massless limit the last equation becomes after introducing a
cutoff $\Lambda$ to regularize the $p_3$ integral
\begin{multline}
n_5
=
\frac{\vert e B \vert \mu_5}{2\pi^2}
\Biggl( 1 + 2
\sum_{n=1}^{ \lfloor \frac{\mu_5^2}{2 \vert e B \vert} \rfloor}
\sqrt{1 - \frac{2 \vert e B \vert n} {\mu_5^2}} \Biggr)
\\
- 
\frac{\vert e B \vert}{2\pi}
\sum_{\pm}
\sum_{s = \pm}
\sum_{n=1}^{\infty}
\int_{-\infty}^{\infty} \frac{\mathrm{d} p_3}{2\pi}
\frac{\mathrm{d} \omega_{p,s}}{\mathrm{d} \mu_5}
 n(\omega_{p,s} \pm \mu)
.
\label{eq:n5magfield}
\end{multline}

\begin{figure}[t]
\includegraphics{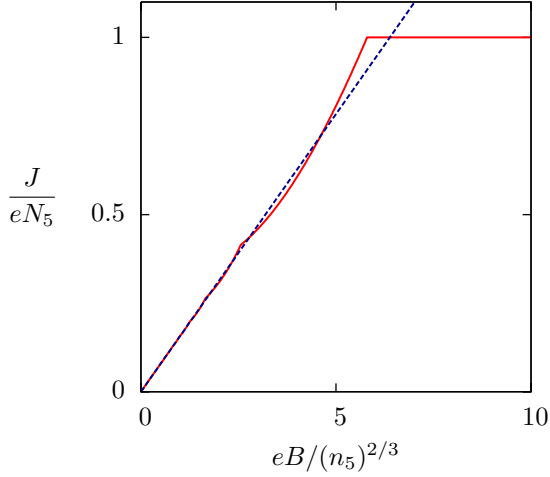}
\caption{Current at zero temperature in a homogeneous magnetic field
as a function of magnetic field strength. 
The dashed line indicates the small field limit approximation.
\label{fig:current}}
\end{figure}

For very large magnetic fields ($\vert e B \vert > \mu_5^2/2$) only
the lowest Landau level (only the first term $1$ in the brackets)
contributes to the current.  Hence $\mu_5 = 2\pi^2 n_5 / \vert e B
\vert$ and the current becomes simply equal to the total chiral charge
in the system,
\begin{equation}
 J =  \mathrm{sgn}(B) \vert e \vert N_5 
\quad \mathrm{if}\quad
 \vert e B \vert > (2\pi^4)^{1/3} n_5^{2/3}.
\label{eq:currentlargefield}
\end{equation}
This result can be easily understood from Eq.~(\ref{eq:thecurdens}).
In a very high magnetic field all modes are fully polarized so that we
have $\langle \phi_{R,L}^\dagger \sigma_3 \phi_{R,L} \rangle = \mathrm{sgn}(eB)
n_{R,L}$.  Applying this to Eq.~(\ref{eq:thecurdens}) gives
Eq.~(\ref{eq:currentlargefield}) \cite{KMW}.
We shall limit our discussions to the $T=\mu=0$ case for a while.  If
$\vert e B \vert < \mu_5^2/2$, not only zeroth but also higher order
Landau levels start to contribute. In Fig.~\ref{fig:current} we have
displayed the current calculated numerically as a function of
$B$.  The current is saturated for $e B/(n_5)^{2/3} > (2\pi^4)^{1/3}
\approx 5.797$.  The small magnetic field limit result
(\ref{eq:zerotempcurrent}) can be written as
\begin{equation}
 J = \frac{ (3 \pi^2)^{1/3}}{2\pi^2} e N_5 \left(\frac{e B}{n_5^{2/3}}\right).
\end{equation}
This limit is displayed as well in Fig.\ref{fig:current} by a dashed line.
The approximation is good as long as the current is not saturated.
The slope is $(3 \pi^2)^{1/3}/ (2\pi^2) \approx 0.1567.$

\begin{figure}[t]
\includegraphics{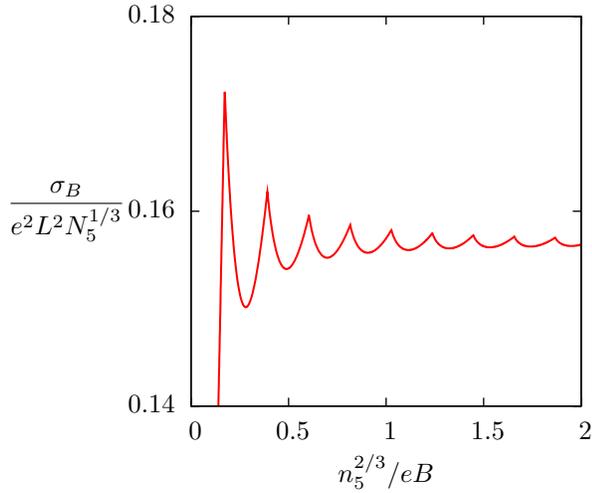}
\caption{Chiral Magnetic conductivity as a function of
the inverse magnetic field strength at zero temperature
and chemical potential.
\label{fig:conduct}} 
\end{figure}

The higher-order Landau levels are creating oscillations in the
conductivity through $n_5$ as can be seen from Fig.~\ref{fig:conduct}
where we have displayed the conductivity versus $n_5^{3/2}/\vert e
B\vert$.  Related oscillations exist in the conductivity induced by an
electric field in the presence of a perpendicular magnetic field. In
that case they are called Shubnikov--de Haas oscillations.  The period
of the oscillations is equal to $\mu_5^2/ 2$ as a function of $\vert
eB \vert$.  Since $\mu_5$ depends on $\vert eB \vert$, the period is
not a constant function of $1 / \vert e B \vert$. In the small
magnetic field limit we can use that $\mu_5 = (3\pi^2)^{1/3}
n_5^{1/3}$.  Hence for small magnetic fields the period of the
oscillations in the conductivity as a function of $n_5^{3/2} / \vert e
B \vert$ becomes $2 / (3\pi^2)^{2/3} \approx 0.2090$.  The mean value
of the Chiral Magnetic conductivity can be found from the small
magnetic field approximation which yields $\sigma_B/(e^2 N_5) = (3
\pi^2)^{1/3}/ (2\pi^2) \approx 0.1567.$

\begin{figure}[t]
\includegraphics{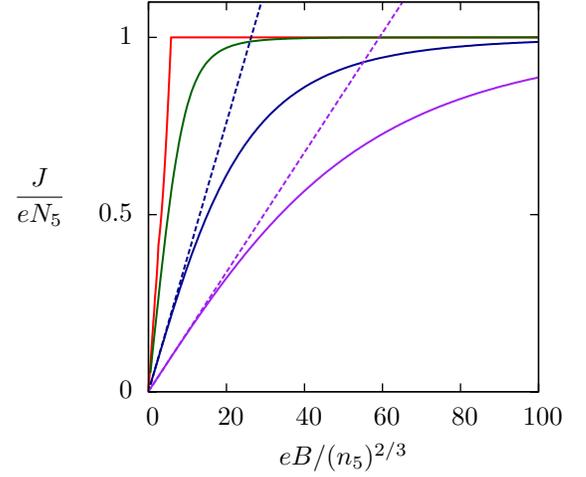}
\caption{Current as a function of magnetic
field for different temperatures. 
Displayed with a solid line are from left to right:
$T/n_5^{1/3} = 0, 1, 2$ and $3$. 
The dashed lines are the small
field approximations for $T/n_5^{1/3} = 2$ and $3$.
\label{fig:curtemp}}
\end{figure}

In order to study the effect of temperature on the current in a
homogeneous magnetic field, we have solved Eq.~(\ref{eq:n5magfield})
numerically. We have displayed the current in Fig.~\ref{fig:curtemp}
for different temperatures. Clearly, at higher temperatures it
requires larger magnetic fields to saturate the current.  This is
because at high temperature more higher momentum modes are occupied,
which are more difficult to polarize.  The dashed lines in
Fig.~\ref{fig:curtemp} denote the small magnetic field approximations
from Eq.~(\ref{eq:hightempcurrent}).  On dimensional grounds one
expects the small magnetic field approximation to be valid for $e B <
T^2$ with $eB < \mu^5$. Indeed this can be seen in the figure, at
finite temperature the small magnetic field approximation is even good
to larger values of the magnetic field than at zero temperature.  The
oscillations in the Chiral Magnetic conductivity will be smeared by
temperature.

\subsection{Implications for heavy-ion collisions}
To obtain the induced current in QCD we can sum the previous results
over flavors and insert the appropriate color factor. We will
give only the high temperature result, since that result will
be relatively insensitive to gluonic corrections and is also the
most relevant for studying the implications of the topological
charge changing transitions in the quark-gluon plasma. Generalizing
Eq.~(\ref{eq:hightempcurrent}) we obtain
\begin{equation}
 J = \frac{3 e^2}{2 \pi^2}
\frac{N_5}{N_f}
\frac{1}{T^2 + \mu^2/\pi^2}
  \langle B \rangle \sum_f q_f^2 .
\label{eq:hightempcurrentqcd}
\end{equation}
Here $q_f e$ is the electric charge carried by quarks of flavor $f$.
Note that $N_f$ in the above is from replacing $\mu_5$ by $N_5$.  This
equation can be used to make predictions for the charge asymmetry in
heavy-ion collisions like is done in Ref.~\cite{KMW}.  In the
quark-gluon plasma we could maybe expect $n_5$ to be several units per
$\mathrm{fm}^3$ deduced from typical QCD sphaleron sizes. In
that case $T/n_5^{1/3} \sim 1 - 10$.  Since $e B \sim
10^4\;\mathrm{MeV}^2$ at the earliest times just after the collision
\cite{KMW}, $e B /n_5^{2/3} \sim 1-10$; it follows from
Fig.~(\ref{fig:curtemp}) that the current is never expected to be
saturated at all. Hence the linear approximation in $B$,
Eq.~(\ref{eq:hightempcurrentqcd}), can be applied to the study of the
Chiral Magnetic Effect in heavy-ion collisions.

In a heavy-ion collision the magnetic field is pointing along the
direction of angular momentum, which is perpendicular to the reaction
plane.  We can define like in Ref.~\cite{KMW}, $\Delta^\pm$ to be the
difference between the total amount of positive/negative charge above
and below the reaction plane in units of $\vert e \vert$. If $\mu$ is
small enough we can assume that the probability to produce a quark is
the same as anti-quark.  Then each time a sphaleron transition with
winding number $Q_\mathrm{w} = -N_5/2 N_f$ is taking place we find that
\begin{eqnarray}
 \Delta^+ &\rightarrow& \Delta^+ \pm \xi_\pm(x_\perp)
\frac{3 \vert Q_\mathrm{w} \vert }{2 \pi^2}
\frac{ \langle e B \rangle}{T^2 + \mu^2/\pi^2}
  \sum_f q_f^2,
\label{eq:deltap}
\\
 \Delta^- &\rightarrow& \Delta^-
\mp 
\xi_\mp(x_\perp)
\frac{3 \vert Q_\mathrm{w} \vert }{2 \pi^2}
\frac{\langle  e B \rangle }{T^2 + \mu^2/\pi^2}
  \sum_f q_f^2.
\label{eq:deltam}
\end{eqnarray}
The $\pm$ and $\mp$ signs in the equations above should be read as
follows. If the winding number $Q_\mathrm{w}$ is negative
(positive), $\Delta^+$ increases (decreases), while $\Delta^-$
decreases (increases). The functions $\xi_\pm(x_\perp)$ defined in
\cite{KMW} are phenomenological screening functions to describe
the effect of the quark-gluon plasma through which the separated
particles have to travel.

The observables proposed in Ref.~\cite{V04} and analyzed in \cite{IVS,
  VQM} are sensitive to the correlators $\langle \Delta^{\pm}
\Delta^\pm\rangle$ and $\langle \Delta^+ \Delta^- \rangle$.  These
correlators can be obtained from Eqs.~(\ref{eq:deltap}) and
(\ref{eq:deltam}) by assuming the one-dimensional random walk picture,
folding it with the sphaleron rate and integrating over time and
volume. This analysis has been performed in Ref.~\cite{KMW}.

In Ref.~\cite{KMW} the current was estimated to be proportional to the
degree of polarization of the quarks with momenta smaller than the
inverse size $1/\rho$ of the typical sphaleron. In Ref.~\cite{KMW}
Eqs.~(\ref{eq:deltap}) and (\ref{eq:deltam}) are similar, except
that the following replacement has to be made,
\begin{equation}
\frac{3}{2\pi^2} \frac{1}{T^2 + \mu^2/\pi^2} \rightarrow 2 \rho^2.
\end{equation}
Since $\rho \approx 1/(\alpha_S T)$, where $\alpha_S$ is the strong
coupling constant, the newly obtained results are slightly different.
The difference stems from the fact that in the calculation in
Ref.~\cite{KMW} the typical momenta were determined by the inverse
size of the sphaleron, $\alpha_s T$, while in the calculation in this
paper, equilibrium was assumed so that the typical momenta are of
order $T$.

\section{Effects of mass and chiral condensate}
In the presence of mass right- and left-handed quarks are coupled. A
chiral condensate does essentially the same. Because the axial charge
density operator does not commute with the Hamiltonian in the presence
of a mass term, chirality is not conserved anymore. Hence the massive
case becomes a dynamical problem and cannot be studied using the
equilibrium approach we used in this article because $\mu_5$
will depend on time when $N_5$ decays.

The effect of mass on the anomaly was studied in Ref.~\cite{AGP}. It
was found that mass term always causes an asymmetry between the number
of right- and left-handed fermions to decay. The time-scale of this
decay depends the typical momentum (the temperature) of the particles,
their mass, and the chiral condensate.  For $T > T_c$ where the
momentum scale is much larger than quark masses the decay time will be
large, so that the equilibrium approach will be reasonably
good. However if the temperature becomes in the neighborhood of $T_c$
the chiral condensate becomes important. Any asymmetry will be washed
out, which will reduce the current.  It would be very interesting to
know how fast the chiral condensate washes out the asymmetry; we will
leave this problem for future study.

\section{The chiral battery}
We would like to point out an interesting hypothetical application of the
Chiral Magnetic Effect -- a rechargeable battery which stores chirality --
the chiral battery.

Let us imagine a hypothetical material with charged fermion
quasi-particles described by the massless Dirac equation.  In
  this Dirac equation the velocity of light is to be replaced by the
  much smaller Fermi velocity $v_F$ of the quasi-particles.  A recent
example of such a material is provided by graphene (for a review see
e.g. \cite{graphene}), even though we should keep in mind that
chirality in graphene is not related to the "usual" spin states
considered above but instead refers to the sub-lattice states. More
directly, our considerations may apply to zero-gap semiconductors with
the linear dispersion relation -- possibly, tellurides.

If we have some finite amount of this material, it can be used as a
battery.  The battery can be charged using the axial anomaly by
placing it in parallel electric and magnetic fields.  The charging
time will be determined by the axial anomaly. The battery stores
energy, since the Fermi-levels of right- and left-handed modes differ.
In a sense, this material is also to be regarded as ``chiral
capacitor.''

In the absence of electric and magnetic fields, chirality is
conserved, so the battery does not discharge.  Now let us connect the
battery to a circuit element with resistance $R$. If we apply a
magnetic field to the battery in the right direction, a current $J$
will be induced due to the Chiral Magnetic Effect.  Note that the
magnetic field alone does no work on fermions in the battery.  The
behavior of this current as a function of the applied magnetic field
and temperature will follow from our analysis in Sec.~III. The current
will cause a potential difference $V = J R$ over the circuit
element. As a result, the same potential difference will also exist
over the battery. Hence an electric field will arise parallel to the
magnetic field.  In this case the axial anomaly operates again to
decrease the chirality. Hence the rate of discharge will be determined
by the axial anomaly as well.

Let us estimate the amount of energy $E$ stored in the chiral battery
per unit volume.  It is equal to the Helmholtz free energy, which is
the energy that can be used to do work. The free energy is the
difference between the thermodynamic potential with a chiral charge
density and without and is easily found by integrating
Eq.~(\ref{eq:n5}) with respect to $\mu_5$. 

We then obtain
\begin{equation}
 E = \Omega(\mu_5) - \Omega(\mu_5=0) 
 =
\frac{1}{12\pi^2}\mu_5^4
+ \frac{1}{6} \mu_5^2 (T^2 + \frac{\mu^2}{\pi^2} ).
\end{equation}
At zero temperature and chemical potential we can use
Eq.~(\ref{eq:n5}) to express $E$ in terms of the chiral charge
density. We find
\begin{equation}
 E = \frac{(3\pi^2)^{1/3}}{4} n_5^{4/3} .
\end{equation}
The typical distance between the lattice sites in a crystal is of
order $0.1\;\mathrm{nm}$.  Suppose we can store $1$ unit of chirality
per lattice site, i.e.\ an excess of 100 right-handed fermions over
left-handed fermions per $\mathrm{nm}^3$. Then the energy density will
be
\begin{equation}
 E = 7.1 \times 10^4 \; \frac{v_F}{c} \frac{\mathrm{eV}}{\mathrm{nm}^3}
   = 1.1 \times 10^7 \; \frac{v_F}{c} \frac{\mathrm{J}}{\mathrm{cm}^3}.
\end{equation}
Here $v_F$ is the Fermi velocity.
 In typical materials like graphene
$v_F / c \sim 10^{-2}$, so the typical storage capacity of the chiral
battery is of order $10^5 \; \mathrm{J}/\mathrm{cm}^3 \simeq
30\;\mathrm{Wh}/\mathrm{cm}^3$.  This is comparable or better
than "conventional" batteries whose energy density is typically
$10-100\;\mathrm{Wh}/\mathrm{Kg}$; note besides that the
current in our case is spin-polarized and so may be used for
spintronic applications.

\section{Conclusions}
A system with a nonzero chirality responds to a magnetic field by
inducing a current along the magnetic field.  This is the Chiral
Magnetic Effect. The behavior of the current as a
function of chirality, baryon chemical potential and temperature has
been obtained in equilibrium in this article. 

The Chiral Magnetic Effect can be studied using heavy-ion collisions.
The possible experimental observation of the Chiral Magnetic Effect
would be direct evidence for the existence and relevance of gluon
configurations with non-trivial topology. Furthermore it will signal
$\mathcal{P}$- and $\mathcal{CP}$-violation in QCD on an
event-by-event basis.  A thorough theoretical understanding of the
Chiral Magnetic Effect will help the experimental analysis by offering
the possibility of more accurate predictions of the observables.

Since the Chiral Magnetic Effect is due to a mixture of QCD and
electromagnetic effects, it has very characteristic behavior.
For example it is expected that the correlators analyzed in experiment
are proportional to the square of the charge of the colliding nuclei.
This very specific behavior can be 
investigated by measuring collisions
of nuclei with the same atomic number but different charge. With
better theoretical understanding, more predictions could be made.

The Chiral Magnetic Effect can only operate in the deconfined, chirally
symmetric phase. Deconfinement is necessary, because quarks need to be
separated over long distances in order for the Chiral Magnetic Effect
to work. Restoration of chiral symmetry is needed, since a chiral
condensate always will wash out any difference between the number of
right- and left-handed quarks. Hence if observed the Chiral
Magnetic Effect might be used as an order parameter for the
confinement/deconfinement and the chiral symmetry
breakdown/restoration transition.

Because the Chiral Magnetic Effect probes the $\mathcal{P}$- and
$\mathcal{CP}$-violating interactions in QCD it can help us to get a
better understanding of the so-called strong $\mathcal{CP}$ problem.
The problem refers to the fact that strong interactions do not break
the $\mathcal{P}$ and $\mathcal{CP}$ symmetries explicitly even though
an addition of $\mathcal{P}$-- and $\mathcal{CP}$--odd $\theta$-term
to the QCD Lagrangian is perfectly allowed without spoiling gauge
invariance. The Chiral Magnetic Effect probes the configurations which
in principle also cause explicit $\mathcal{P}$- and $\mathcal{CP}$
violation if $\theta$ is non-vanishing.

The Chiral Magnetic Effect has also a nice analogy in the physics of
the Early Universe. One mechanism to explain the matter-antimatter
asymmetry is electroweak baryogenesis \cite{KRS,
  Shaposhnikov1987}. There electroweak sphalerons induce via the axial
anomaly $\mathcal{C}$- and $\mathcal{CP}$-odd effects. As a result
baryon plus lepton number is generated. This process is very similar
to the Chiral Magnetic Effect.  It is also quite possible that the
Chiral Magnetic Effect itself could have an important role in the
Early Universe if a large magnetic field and/or a non-zero expectation
value of the axion field were present at that time.

\section*{Acknowledgments}
We are grateful to Larry McLerran and Eric Zhitnitsky for discussions; we thank Larry McLerran also for comments on the
manuscript.

This manuscript has been authored under Contract
No.~\#DE-AC02-98CH10886 with the U.S.\ Department of Energy.  K.~F.\
is supported by Japanese MEXT grant no.\ 20740134 and also supported
in part by Yukawa International Program for Quark Hadron Sciences.

\appendix

\end{document}